\documentclass[superscriptaddress,showkeys,useAMS,twocolumn,nofootinbib]{revtex4-1}
\pdfoutput=1
\usepackage{graphics, graphicx, amsmath, amssymb}
\usepackage{epstopdf}
\usepackage{dsfont}
\usepackage{hyperref}
\usepackage{color}
\usepackage[normalem]{ulem}

\newcommand{\eq}[1]{\begin{equation}\begin{aligned}#1\end{aligned}\end{equation}}
\newcommand{\iu}{\text{i}}
\newcommand{\eu}{\text{e}}
\newcommand{\ha}{\hat{a}}
\newcommand{\had}{\hat{a}^\dagger}
\newcommand{\hb}{\hat{b}}
\newcommand{\hbd}{\hat{b}^\dagger}
\newcommand{\ket}[1]{\left|#1\right\rangle}
\newcommand{\bra}[1]{\left\langle#1\right|}
\newcommand{\braket}[2]{\left.\left\langle#1\right|#2\right\rangle}
\newcommand{\expct}[1]{\left\langle#1\right\rangle}

\newcommand{\textblue}[1]{{\color{black}#1}}

\begin{document}
\title{Nonclassical states that generate zero entanglement with a beam splitter}
\date{\today}
\author{Aaron Z. Goldberg}
\email{goldberg@physics.utoronto.ca}
\affiliation{Department of Physics, University of Toronto, Toronto, ON, M5S 1A7}
\author{Daniel F. V. James}
\affiliation{Department of Physics, University of Toronto, Toronto, ON, M5S 1A7}
\begin{abstract}
	Beam splitters are routinely used for generating entanglement. Their entangling properties have been studied extensively, with nonclassicality of the input states a prerequisite for entanglement at the output. Here we quantify the amount of entanglement generated by weakly-reflecting beam splitters, and look for nonclassical states that are not entangled by general beam splitters. We extend the known class of results to mixed and non-Gaussian states, finding that inputting highly nonclassical combinations of unpolarized states that are squeezed and displaced onto a beam splitter can still yield separable output states. This is important in light of the challenge of characterizing mixed state entanglement. We further identify a parallel between SU(2) unpolarized states and the two-mode vacuum. Our result is crucial for understanding the generation of modal entanglement by beam splitters.
\end{abstract}

\maketitle

\section{Introduction}
Entanglement is a hallmark of quantum mechanics. It is responsible for the superiority of many quantum communication \cite{BennettWiesner1992,Bennettetal1993,Bouwmeesteretal1997,Renetal2017} and computation \cite{Shor1995,ChuangYamamoto1995,DiVincenzo2000} protocols over their classical counterparts, and has fascinated physicists for decades \cite{EPR1935,Schrodinger1935cat,Schrodinger1935,Bell1964,Bell1966,Aspectetal1981,Hensenetal2015,Shalmetal2015,Giustinaetal2015}. 
One of the most common routines for generating entanglement is by using beam splitters.

The entangling properties of beam splitters have generated much theoretical interest \cite{Tanetal1991,Sanders1992,HuangAgarwal1994,Paris1999,Kimetal2002,Xiangbin2002,Xiangbin2002Gaussian,Wolfetal2003,Asbothetal2005,Tahiraetal2009,Zhangetal2013,Killoranetal2014,VogelSperling2014,Geetal2015,Killoranetal2016,GholipourShahandeh2016}. A single photon incident on a beam splitter can yield a nonlocal, entangled state \cite{Tanetal1991,vanEnk2005}, as can a superposition of coherent states \cite{Sanders1992}. Nonclassical input states such as squeezed states typically generate entangled outputs \cite{HuangAgarwal1994,Paris1999,Xiangbin2002Gaussian,Wolfetal2003}, with the {known pure state} exceptions {being} a particular class of squeezed states \cite{HuangAgarwal1994,Zhangetal2013}. It was shown that no classical input states incident on a beam splitter can produce entangled outputs \cite{Kimetal2002,Xiangbin2002}; however, there is no guarantee that nonclassical input states will indeed generate entanglement \cite{HuangAgarwal1994,SperlingVogel2009negativequasi,Zhangetal2013}.  
Here we investigate a more general condition for incident states to yield separable, as opposed to entangled, outputs{, without restricting ourselves to Gaussian nor pure states}.

\textblue{When a state is incident on one port of a beam splitter with the vacuum in the other port, nonclassicality of the first state is a necessary and sufficient condition for entanglement generation \cite{Asbothetal2005}. This has lead to some confusing language by various authors, who occasionally write without qualification that any nonclassicality in the input modes guarantees entanglement at the output; see, for example, key phrases in Refs. \cite{Tahiraetal2009,Tejetal2013,Killoranetal2014,DeyHussin2015}. The necessary and sufficient conditions for a general, separable two-mode state to produce entanglement at a beam splitter have not yet been elucidated, and are the subject of our current investigation.}

Characterizing entanglement for mixed states is an active area of research. 
Separable states have a Werner decomposition \cite{Werner1989} and a positive partial transpose \cite{Peres1996,Horodeckietal1996}, satisfying a set of Bell-like inequalities \cite{Terhal2000,Duanetal2000}. Entangled states, on the other hand, always have entanglement witnesses \cite{Horodeckietal1996}, and the entanglement properties of infinite-dimensional systems can be fully characterized in finite-dimensional subspaces \cite{SperlingVogel2009finite}. Moreover, finite-dimensional states can always be decomposed into a sum of two separable mixed states, albeit with non-positive-definite coefficients \cite{VidalTarrach1999}. Nonetheless, readily-computable necessary and sufficient conditions for entanglement exist only for $2\times 2$ and $3\times 3$ density matrices \cite{Horodeckietal1996} and for Gaussian states \cite{SperlingVogel2009finite}; determining whether a given state is entangled is NP-hard \cite{Gurvits2004}. It thus remains a relevant question to find classes of {non-Gaussian and mixed} states that do and do not generate entanglement through the action of a beam splitter.

{In this paper we address the challenge of characterizing mixed state entanglement by presenting the first study of non-Gaussian mixed states that remain separable at a beam splitter.}
To do this we investigate properties of the Gaussian pure and mixed states found by Refs. \cite{HuangAgarwal1994,Zhangetal2013} to remain separable at the output of a beam splitter; we extend Ref. \cite{Zhangetal2013}'s $n$-mode pure state result to mixed states, and Ref. \cite{HuangAgarwal1994}'s two-mode Gaussian mixed state result to non-Gaussian mixed states \cite{PrakashChandra1971,Agarwal1971}.

\textblue{Our key insight into this problem makes use of the SU(2) unpolarized states [also known as SU(2)-invariant states, type I unpolarized light, and quantum unpolarized states; see Refs. \cite{PrakashChandra1971,Agarwal1971,Lehneretal1996,Bjorketal2010,Klimovetal2005}]. SU(2) unpolarized states are states whose polarization vectors have zero length, and whose polarization moments are isotropic to all orders \cite{PrakashChandra1971,Agarwal1971,Lehneretal1996}. SU(2) unpolarized states are in some senses ``the opposite" of classical states \cite{Zimba2006,Martinetal2010,Crannetal2010,Bjorketal2015,BaguetteMartin2017}; this is in contrast with states that are perfectly polarized, which have polarization vectors of maximal length and have all of their quanta being in the same mode \cite{GoldbergJames2017}. We show that the SU(2) unpolarized states are separable, and act like the vacuum in that they emerge unaltered from arbitrary beam splitters. We use these states to construct sets of non-Gaussian mixed states that remain separable at beam splitters.} It is striking that the SU(2) unpolarized states,  \cite{Zimba2006,Martinetal2010,Crannetal2010,Bjorketal2015,BaguetteMartin2017} \textblue{which are highly nonclassical}, lie at the heart of the states that do not generate entanglement via beam splitters.

\section{Beam splitters, SU(2) operations, and entanglement}
We start by considering two orthogonal bosonic modes associated with the annihilation operators $\ha$ and $\hb$, such as two spatial or polarization modes of light.
The action of a beam splitter with reflectivity $\sin^2\frac{\theta}{2}$ on the two modes is represented by an SU(2) operator \cite{MandelWolf1995}
\eq{\hat{R}\equiv\hat{R}\left(\theta,\varphi\right)&=\exp\left(-\xi\had\hb+\xi^*\hbd\ha\right),\quad\xi=\frac{\theta}{2}\eu^{-\iu\varphi},}
which generates the transformation
\eq{\hat{R}
	\begin{pmatrix}
		\ha\\\hb
	\end{pmatrix}\hat{R}^\dagger=\begin{pmatrix}
	\cos\frac{\theta}{2}&\eu^{-\iu\varphi}\sin\frac{\theta}{2}\\
	-\eu^{\iu\varphi}\sin\frac{\theta}{2}&\cos\frac{\theta}{2}
\end{pmatrix}
\begin{pmatrix}
	\ha\\\hb
\end{pmatrix}.
\label{eq:beam splitter matrix transformation}
}
The beam splitter acting on an $N$-particle state with all particles in one mode $\ket{N}\otimes\ket{0}\propto\had\vphantom{a}^N\ket{\text{vac}}$ rotates the state from the north pole of the Poincar\'e sphere to the angular coordinates $\left(\theta,\varphi\right)$: \eq{
	\ket{\theta\varphi^{(N)}}&\equiv \hat{R}\ket{N}\otimes\ket{0}\\
	&=\sum_{m=0}^Nc_m^{(\theta,\varphi)}\ket{m}\otimes\ket{N-m}
	\label{eq:expanded R psi},
}
where the coefficients are given by 
\eq{c_m^{(\theta,\varphi)}=\sqrt{\binom{N}{m}}\cos^m\frac{\theta}{2}\sin^{N-m}\frac{\theta}{2}\eu^{\iu\varphi\left(N-m\right)}.}

All two-mode pure states can be rewritten using a Schmidt decomposition into the form \eq{\ket{\Psi}=\sum_kv_k\ket{\psi_k}\otimes\ket{\phi_k}.} The number of nonzero coefficients $v_k$ is known as the Schmidt rank, and a pure state is separable if and only if its Schmidt rank is 1 \cite{NielsenChuang2000}. The state given by Eq. (\ref{eq:expanded R psi}) has Schmidt number $N+1$ and is therefore entangled (for $N>0$ and $\theta\neq 0,\pi$).

Similarly, since SU(2) operations preserve particle number,
\eq{\hat{R}\left(\sum_Nc_N\ket{N}\right)\otimes\ket{0}=\sum_Nc_N\ket{\theta\varphi^{(N)}}.} Except for special cases, the resulting state exhibits entanglement between modes $\ha$ and $\hb$ even though it is perfectly polarized \cite{GoldbergJames2017}; a general pure state input with the vacuum on a beam splitter yields an entangled output. We here seek sets of input states that are \textit{not} entangled by linear optical devices.

\section{Entanglement generated by rotations of pure states}

We first ask how much modal entanglement is generated by a small rotation angle $\theta$ of a general pure state, corresponding to a weakly-reflecting beam splitter. 
We quantify entanglement using the purity of the partial trace with respect to one of the Hilbert spaces 
\eq{\mathcal{E}_\text{p}=1-\text{Tr}_b\left[\hat{\rho}_\text{b}^2\right], \label{eq:entanglement purity}}
 where
$\hat{\rho}_b=\text{Tr}_a\left[\hat{\rho}\right].$ This conveys the information from the Schmidt decomposition through the relation $\mathcal{E}_\text{p}=1-\sum_kv_k^2$. All states have $0\leq \mathcal{E}_\text{p}\leq 1$, with fully separable states satisfying $\mathcal{E}_\text{p}=0$; the deviation from zero is a measure of the entanglement \cite{Barnumetal2003,Barnumetal2004}.

\subsection{Perfectly polarized input}
\label{sec:perfectly polarized input}
We begin in the separable, perfectly polarized \cite{GoldbergJames2017} state $\ket{\Psi}=\ket{\psi}\otimes\ket{0}$, where $\ket{\psi}=\sum_Nc_N\ket{N}$ and $\sum_N \left|c_N\right|^2=1$. 
Since the initial state starts at the north pole of the Poincar\'{e} sphere, azimuthal symmetry allows us to choose $\varphi=0$.
To order $\mathcal{O}\left(\theta^2\right)$, $\hat{R}\ket{\Psi}\approx \ket{\Psi}+\frac{\theta}{2}\ha\hbd\ket{\Psi}+\frac{\theta^2}{8}\ha^2\hbd\vphantom{a}^2\ket{\Psi}-\frac{\theta^2}{8}\had\ha\ket{\Psi}$ (using $\hb\ket{\Psi}=0$). 
We calculate, to the same order,
\eq{
	\hat{\rho}_b&=\text{Tr}\left[\hat{R}\ket{\Psi}\bra{\Psi}\hat{R}^\dagger\right]\\
	&\approx \ket{0}\bra{0}+\frac{\theta}{2}\left(\expct{\ha}\ket{1}\bra{0}+\expct{\had}\ket{0}\bra{1}\right)\\
	&+\frac{\theta^2}{4}\expct{\had\ha}\left(\ket{1}\bra{1}-\ket{0}\bra{0}\right)\\
	&+\frac{\theta^2}{8}\left(\expct{\ha^2}\ket{2}\bra{0}+\expct{\had\vphantom{a}^2}\ket{0}\bra{2}\right)
}
where expectation values are taken with respect to $\ket{\psi}$ and the operators are in the Hilbert space associated with $\hb$. Then we find
\eq{
	\mathcal{E}_\text{p}&=\frac{\theta^2}{2}\left(\expct{\had\ha}-\expct{\had}\expct{\ha}\right)+\mathcal{O}\left(\theta^4\right)
	\label{eq:order theta2 entanglement}.
}

The amount of entanglement from a small rotation is seen to grow with the square of the rotation angle and with the average number of particles initially in mode $\ha$. It decreases when the particles are initially more evenly distributed between the states with different particle number. 

Defining $\ket{\tilde{\psi}}=\ha\ket{\psi}$, we have, by the Cauchy-Schwartz inequality,
\eq{
	\braket{\psi}{\psi}\braket{\tilde{\psi}}{\tilde{\psi}}&\geq \left|\braket{\psi}{\tilde{\psi}}\right|^2 \\
	\braket{\tilde{\psi}}{\tilde{\psi}}&\geq \left|\braket{\psi}{\tilde{\psi}}\right|^2\\
	\bra{\psi}\had\ha\ket{\psi}&\geq\bra{\psi}\had\ket{\psi}\bra{\psi}\ha\ket{\psi}.
}
This means that the $\mathcal{O}\left(\theta^2\right)$ term in Eq. (\ref{eq:order theta2 entanglement}) only vanishes when the Cauchy-Schwartz inequality is saturated. The inequality is saturated if and only if $\ket{\tilde{\psi}}$ is linearly dependent on $\ket{\psi}$; i.e., only $\ha\ket{\psi}\propto \ket{\psi}$ can remain unentangled to lowest order in $\theta$. Thus a coherent state $\ket{\psi}=\ket{\alpha}\equiv\exp\left(\alpha\had-\alpha^*\ha\right)\ket{\text{vac}}$ is the only {single-mode pure} state that can remain unentangled following the action of the rotation operator;
every pure state that is not a coherent state, input on a beam splitter with the vacuum in the other port, generates entanglement at the output.

\subsection{General two mode state}
Next we generalize to the case of inputting an arbitrary, separable pure state $\ket{\Psi}=\ket{\psi}\otimes\ket{\phi}$ on the beam splitter. In this case it is no longer possible to assume $\varphi=0$.
We note that for coherent states input on both modes, Eq. (\ref{eq:beam splitter matrix transformation}) always yields a separable output

\eq{\hat{R}\left(\theta,\varphi\right)\ket{\alpha}\otimes\ket{\beta}=&\ket{\alpha\cos\frac{\theta}{2}-\beta\eu^{-\iu\varphi}\sin\frac{\theta}{2}} \\&\quad\otimes\ket{\alpha\eu^{\iu\varphi}\sin\frac{\theta}{2}+\beta\cos\frac{\theta}{2}},
	\label{eq:rotation on coherent state}
}
which agrees with the above results when $\beta=0$.

For the general two mode case, we show in Appendix \ref{app:rotating two mode state} that
\eq{
	\mathcal{E}_\text{p}\approx \theta^2\left(AB+\frac{A+B}{2}-\Re\left[\eu^{2\iu\varphi}\Delta^2\hbd\Delta^2\ha\right]\right),
}
where
\eq{
	&A\equiv \expct{\had\ha}-\expct{\had}\expct{\ha},\\
	&B\equiv \expct{\hbd\hb}-\expct{\hbd}\expct{\hb},\\
	&\Delta^2\hat{O}=\expct{\hat{O}^2}-\expct{\hat{O}}^2,
}
and  expectation values are taken with respect to $\ket{\Psi}$. As before, $A,B\geq 0$, with the equalities holding only for modes $\ha$ and $\hb$ being coherent states. Again, entanglement grows as $\theta^2$. We have thus shown that increasing the reflectivity of a \textblue{weakly-reflective} beam splitter increases the amount of entanglement that it produces.

As a corollary, 
we mention that this general two mode scenario has the additional possibility of retaining separability if $\Re\left[\eu^{2\iu\varphi}\Delta^2\hbd\Delta^2\ha\right]=AB+\frac{A+B}{2}$. This condition has been shown to be true for the product of two single-mode squeezed states that are squeezed by the same amount \cite{Zhangetal2013}; i.e., the state $\ket{\Psi}=\hat{S}_a\left(\gamma\right)\otimes\hat{S}_b\left(\eu^{-2\iu\varphi}\gamma\right)\ket{\text{vac}}$ and displacements thereof, for squeeze operator $\hat{S}_c\left(\gamma\right)=\exp\left[\left(\gamma\hat{c}^2-\gamma^*\hat{c}^\dagger\vphantom{a}^2\right)/2\right]$, remain separable after going through a beam splitter. Ref. \cite{Zhangetal2013} showed that the states \eq{
	\ket{\Phi}=\hat{D}_a\left(\alpha\right)\hat{S}_a\left(\gamma\right)\otimes \hat{D}_b\left(\beta\right)\hat{S}_b\left(\eu^{-2\iu\varphi}\gamma\right)\ket{\text{vac}}
	\label{eq:non classical pure states that do not generate entanglement}
}
are the only pure states that remain unentangled when input on a beam splitter [for displacement operators satisfying $\hat{D}_a\left(\alpha\right)\otimes\hat{D}_b\left(\beta\right)\ket{\text{vac}}=\ket{\alpha}\otimes\ket{\beta}$;  {we have been more explicit than Ref. \cite{Zhangetal2013} in allowing for any beam splitter phase}]. 
We note that this condition only holds when the relative phase of the squeezing between the two modes takes a value appropriate to the phase imparted by the beam splitter. Even though these states are highly nonclassical, they generate zero entanglement at a beam splitter. 

\textblue{This small-$\theta$ treatment of beam splitters was sufficient for finding the pure states that generate no entanglement at beam splitters. Unfortunately, we were unsuccessful at applying such a treatment to the general mixed state case. We proceed to the mixed state case using an alternative method.}

\section{Effect of rotation on mixed states}
\subsection{Mixed state entanglement}
We now generalize the above results to mixed states.\footnote{{We are specifically investigating mixed states that are not pure, confronting the challenge of characterizing mixed state entanglement in this scenario.}}
The separability condition for mixed states is that they can be written in the form
\eq{
	\hat{\rho}=\sum_k p_k\hat{\rho}_k^{(A)}\otimes \hat{\rho}_k^{(B)},
	\label{eq:werner decomposition separable mixed states}
}
where $\sum_k p_k=1$, $0< p_k\leq 1$, and $\hat{\rho}_k^{(A)}$ and $\hat{\rho}_k^{(B)}$ are density operators \cite{Werner1989}.
We know by Eq. (\ref{eq:rotation on coherent state}) that any state of the form 
\eq{
	\hat{\rho}_0=\sum_k p_k\ket{\alpha_k}\bra{\alpha_k}\otimes\ket{\beta_k}\bra{\beta_k}
	\label{eq:rho made from diagonal sum of coherent states}
}
transforms as
\begin{widetext}
\eq{
	\hat{R}\rho_0\hat{R}^\dagger=\sum_k p_k& \ket{\alpha\cos\frac{\theta}{2}-\beta\eu^{-\iu\varphi}\sin\frac{\theta}{2}}\bra{\alpha\cos\frac{\theta}{2}-\beta\eu^{-\iu\varphi}\sin\frac{\theta}{2}}
 	\otimes\ket{\alpha\eu^{\iu\varphi}\sin\frac{\theta}{2}+\beta\cos\frac{\theta}{2}}\bra{\alpha\eu^{\iu\varphi}\sin\frac{\theta}{2}+\beta\cos\frac{\theta}{2}}.
}
\end{widetext}
This satisfies the separability condition, so every state of the form of Eq. (\ref{eq:rho made from diagonal sum of coherent states}) remains separable after being rotated. In fact, due to the necessity to only sample finite-dimensional subspaces to verify entanglement \cite{SperlingVogel2009finite}, any state $\hat{\rho}_0=\int \text{d}^2\alpha\, \text{d}^2\beta\, P\left(\alpha,\beta\right)\ket{\alpha}\bra{\alpha}\otimes\ket{\beta}\bra{\beta}$ with $P\left(\alpha,\beta\right)\geq 0$ will remain separable after being rotated by a beam splitter \cite{Kimetal2002,Xiangbin2002}. 

Only nonclassical states, in the sense of having \textblue{negative} Glauber-Sudarshan $P$-functions \textblue{\cite{Glauber1963,Sperling2016}},\footnote{\textblue{States with Glauber-Sudarshan $P$-functions that are more singular than delta functions, but do not require infinitely many derivatives of delta functions, are also guaranteed to be nonclassical \cite{Sperling2016}.}} can generate entanglement. This has led \textblue{some} authors to mistakenly \textblue{assert} that all nonclassical states will generate entanglement;
whereas, nonclassicality is a necessary but not sufficient condition. In truth, one must show that \textit{no} quasiprobability distributions for a given state are positive definite to guarantee that the state will generate entanglement, an analytically intractable problem \cite{Gurvits2004,SperlingVogel2009negativequasi}. We thus address the relevant question of finding {mixed} states that are nonclassical, yet still do not generate entanglement.

\subsection{Set of nonclassical mixed states that do not generate entanglement}
\label{sec:mixed states no entanglement}
We first examine the properties of the states given by Eq. (\ref{eq:non classical pure states that do not generate entanglement}) that allow them to remain separable under the action of a beam splitter. For a general product of squeeze operators, Eq. (\ref{eq:beam splitter matrix transformation}) implies that
\eq{
	\hat{R}\hat{S}_a\left(\gamma_a\right)&\otimes\hat{S}_b\left(\gamma_b\right)\hat{R}^\dagger=\\
	&\exp\left[\frac{\ha^2}{2}\left(\gamma_a\cos^2\frac{\theta}{2}+\gamma_b\eu^{2\iu\varphi}\sin^2\frac{\theta}{2}\right) \right.\\
	&\quad+ \frac{\hb^2}{2}\left(\gamma_b\cos^2\frac{\theta}{2}+\gamma_a\eu^{-2\iu\varphi}\sin^2\frac{\theta}{2}\right) \\
	&\quad\left.+\frac{\ha\hb}{2}\sin\theta\left(\gamma_a\eu^{-\iu\varphi}-\gamma_b\eu^{\iu\varphi}\right)
	-\text{H.c.}\right].
	\label{eq:squeeze operators unaffected by beam splitter}
}
We immediately find that $\hat{R}\hat{S}_a\left(\gamma_a\right)\otimes\hat{S}_b\left(\gamma_b\right)\hat{R}^\dagger=\hat{S}_a\left(\gamma_a\right)\otimes\hat{S}_b\left(\gamma_b\right)$ for $\gamma_a=\eu^{2\iu\varphi}\gamma_b$.
An SU(2) rotation of the state in Eq. (\ref{eq:non classical pure states that do not generate entanglement}) yields
\eq{
	\hat{R}\ket{\Phi}
	&=\hat{\mathbf{D}}_a\left(\alpha\right)\hat{\mathbf{S}}_a\otimes\hat{\mathbf{D}}_b\left(\beta\right)\left(\gamma\right)\hat{\mathbf{S}}_b\left(\eu^{-2\iu\varphi}\gamma\right)\hat{R}\ket{\text{vac}},
}
where $\hat{\mathbf{A}}=\hat{R}\hat{A}\hat{R}^\dagger$ and we have used $\hat{R}^\dagger\hat{R}=\mathds{I}$. Eq. (\ref{eq:beam splitter matrix transformation}) can be applied to all of the creation and annihilation operators inside of the displacement operators, in addition to applying Eq. (\ref{eq:squeeze operators unaffected by beam splitter}), to yield the manifestly separable state
\eq{
	\hat{R}\ket{\Phi}=&\hat{D}_a\left(\alpha\cos\frac{\theta}{2}-\beta\eu^{-\iu\varphi}\sin\frac{\theta}{2}\right)\hat{S}_a\left(\gamma\right)\\
	&\otimes\hat{D}_b\left(\beta\cos\frac{\theta}{2}+\alpha\eu^{\iu\varphi}\sin\frac{\theta}{2}\right)
	\hat{S}_b\left(\eu^{-2\iu\varphi}\gamma\right)\ket{\text{vac}},
}
where we have also used the property $\hat{R}\ket{\text{vac}}=\ket{\text{vac}}$. The set of density operators formed by the pure states given by Eq. (\ref{eq:non classical pure states that do not generate entanglement}) form a convex subspace because the set of separable density operators is convex\textblue{. This leads to our result that} the general density operators 
\eq{
	\hat{\rho}&=\int \text{d}k\, g\left(k\right)\ket{\Phi\left(k\right)}\bra{\Phi\left(k\right)},\quad g\left(k\right)\geq 0\\
	&\quad\quad 	\quad\quad\ket{\Phi\left(k\right)}=\hat{T}\left(k\right)\ket{\text{vac}}\\
	\hat{T}\left(k\right)&=\hat{D}_a\left[\alpha\left(k\right)\right]\hat{S}_a\left[\gamma\left(k\right)\right]
	\otimes\hat{D}_b\left[\beta\left(k\right)\right]\hat{S}_b\left[\eu^{-2\iu\varphi}\gamma\left(k\right)\right]
}
remain unentangled following an SU(2) rotation.

The salient properties of the above states are that (i) rotations of the displacement operators yield new displacement operators acting on the original modes, (ii) rotations of the squeeze operators along the correct axes do not change the operators, and (iii) $\hat{R}\ket{\text{vac}}\bra{\text{vac}}\hat{R}^\dagger=\ket{\text{vac}}\bra{\text{vac}}$, where the latter is a separable state $\ket{0}\bra{0}\otimes\ket{0}\bra{0}$. We can extend our set of states that do not generate entanglement by considering more states that adhere to condition (iii). These are related to studies of unpolarized light.

A new definition for unpolarized light was proposed in 1971, defined as the set of states that are unaffected by SU(2) operations \cite{PrakashChandra1971,Agarwal1971}. The authors found that, since SU(2) operations conserve particle number, the most general state that remains unchanged by an SU(2) operation is the state
\eq{
	\label{eq:unpolarized state}
	\hat{\rho}_\text{un}\left(k\right)=\sum_N\lambda_N\left(k\right)\hat{\mathds{I}}_N,\quad\sum_N\lambda_N\left(k\right)\left(N+1\right)=1,
}
where $\hat{\mathds{I}}_N=\sum_{m=0}^{N}\ket{m}\bra{m}\otimes\ket{N-m}\bra{N-m}$ is the projection operator onto the $N$-particle basis \cite{PrakashChandra1971,Agarwal1971}. The SU(2) unpolarized states, lying at the centre of the Poincar\'{e} sphere, are the opposite of the SU(2) coherent states, which lie on the Poincar\'{e} sphere's surface \cite{Zimba2006}. Since SU(2) coherent states are in some sense the most classical spin states, many authors regard SU(2) unpolarized states as epitomizing ``quantumness" \cite{Klyshko1992,Zimba2006,Martinetal2010,Crannetal2010,Bjorketal2015,BaguetteMartin2017}. Nonetheless, we show in Appendix \ref{app:unpolarized states separable} that the SU(2) unpolarized states are all separable, which allows us to extend condition (iii) above. We also note \textblue{in Appendix \ref{app:unpolarized states separable}} that these states include the special case of two-mode thermal states with each mode at the same temperature.

Since $\hat{\rho}_\text{un}$ is a separable state that remains unchanged under the action $\hat{R}\hat{\rho}_\text{un}\hat{R}^\dagger$, it immediately follows that the density operators
\eq{
	\label{eq:unpolarized state squeezed displaced}
	\hat{\rho}\left(k\right)&=	
	\hat{T}\left(k\right)\hat{\rho}_\text{un}\left(k\right)\hat{T}^\dagger\left(k\right)
}
remain unentangled following SU(2) rotations.
These again form a convex set, and so we have our general result for a set two-mode of mixed states that remain separable under the action of passive linear optics:
\eq{
	\hat{\rho}&=\int \text{d}k\, g\left(k\right)
	\hat{\rho}\left(k\right),\quad g\left(k\right)\geq 0.
	\label{eq:general no entanglement generation}
}
These are the states obtained by squeezing unpolarized light along certain axes and arbitrarily displacing the results. They include the Gaussian states found by Ref. \cite{HuangAgarwal1994} and the pure states found by Ref. \cite{Zhangetal2013}, as well as {mixed and non-Gaussian} generalizations thereof. It is indeed remarkable that, although most nonclassical states generate entanglement at a beam splitter, there exist highly nonclassical states that generate no entanglement in the same situation.

\textblue{\section{Properties of the nonclassical states that generate no entanglement}}
\textblue{We have shown that the states given by Eq. (\ref{eq:general no entanglement generation}) generate no entanglement at beam splitters. We here discuss some important properties of these states.}

As a basic example \textblue{of the states that generate no entanglement}, we see that combinations of displaced number states can remain separable at a beam splitter. For example, the states
	\eq{
\textblue{		\hat{\rho}^{(N)}=\frac{1}{N+1}\sum_{m=0}^{N}\ket{m,\alpha}\bra{m,\alpha}\otimes\ket{N-m,\beta}\bra{N-m,\beta}}
	}
remain separable at \textit{any} beam splitter regardless of the beam splitter's phase $\varphi$ \textblue{and reflectivity parameter $\theta$}, where we have defined displaced number states by $\ket{n,\alpha}=\hat{D}\left(\alpha \right)\ket{n}$. These states are highly impure and non-Gaussian for $N>0$, yet still generate zero entanglement. 

\textblue{To generate an SU(2) unpolarized state, the easiest method is to use natural or thermal light (Appendix \ref{app:unpolarized states separable}) \cite{Lehneretal1996}. Alternatively, one can consider the average output from a laser that generates light without any polarization dependence \cite{Lehneretal1996}. The general state that is output from a single-mode laser can be expressed as
\eq{\hat{\rho}_\text{L}=\frac{\eu^{-\left|\alpha\right|^2}}{2\pi} \sum_N\frac{\left|\alpha\right|^{2N}}{N!}\ket{N}\bra{N}.}
 If the output polarization fluctuates randomly, the output is then 
 an SU(2) unpolarized state \cite{Lehneretal1996}:
\eq{
\hat{\rho}_\text{L,avg}
=\sum_N\frac{\eu^{-\left|\alpha\right|^2}\left|\alpha\right|^{2N}}{\left(N+1\right)!}\hat{\mathds{I}}_N.
}
These states can then be displaced and squeezed using standard optical elements to create more general SU(2) unpolarized states.

Only states with nonclassical $P$-distributions can be used to generate entanglement. Since the set of states given by Eq. (\ref{eq:general no entanglement generation}) are $P$-nonclassical yet generate no entanglement, perhaps there exists a more restrictive type of nonclassicality that is necessary for entanglement generation. It is thus useful to further characterize the nonclassicality present in our states.

The SU(2) unpolarized states fall into the subclass of separable states known as classical-quantum separable states. These are states that can be expressed as $\hat{\rho}=\sum_m p_m \ket{m}\bra{m}\otimes\hat{\rho}_m$ for orthonormal states $\left\{\ket{m}\right\}$ and coefficients satisfying $p_m>0,\,\sum_mp_m=1$ \cite{Dakicetal2010}. It follows that the SU(2) unpolarized states, like all classical-quantum correlated states, have no quantum discord. Quantum discord is a measure of the nonclassical correlations present in a mixed state, and is more general than quantum entanglement \cite{Dakicetal2010,LangCaves2010}. Zero discord could therefore be a necessary condition for mixed states to generate separable states. 

Local unitaries do not affect  discord \cite{GiordaParis2010}, making the states given by Eq. (\ref{eq:unpolarized state squeezed displaced}) also discord-free. However, unlike separable states, the set of states with zero discord is not convex \cite{LangCaves2010}. In general, states given by Eq. (\ref{eq:general no entanglement generation}) have nonzero discord, yet they still generate zero entanglement; this nonclassicality feature can be present in states that generate no entanglement. It follows that states with both zero and nonzero discord can generate both entangled and separable states, as is also clear from the pure state scenario, in which states are separable if and only if they have zero discord. This presently relegates the zero-discord property of the states given by Eq. (\ref{eq:unpolarized state squeezed displaced}) to a mere curiosity.

Another method of seeing the nonclassicality of the states that generate no entanglement is by looking more closely at the states' $P$-distributions. If these distributions are negative, or have finitely many derivatives of delta functions, then the states that they represent are nonclassical \cite{Sperling2016}. Importantly, one can always find a filtered $P$-distribution that is finite everywhere yet negative somewhere for any nonclassical state \cite{KieselVogel2010,Agudeloetal2013}.

It is straightforward to see that the states that generate no entanglement are $P$-nonclassical. For simplicity, one can consider the state given by Eq. (\ref{eq:unpolarized state}) for which a single coefficient $\lambda_N=1/(N+1)$ is nonzero. Indeed, since the $P$-distribution for a state $\ket{m}\bra{m}$ is given by $P\left(\alpha,\alpha^*\right)\propto \frac{\partial^{2m}}{m!\partial\alpha^m\partial\alpha^{*m}}\delta^{(2)}\left(\alpha\right)$ \cite{AgarwalTara1991}, a finite convex combination of Fock states with finite $N$ yields a $P$-distribution made from a finite sum of finitely-many derivatives of delta functions:
\eq{
	P\left(\alpha,\alpha^*;\beta,\beta^*\right)\propto\\ \sum_{m=0}^N\binom{N}{m}\frac{\partial^{2N}}{\partial\alpha^m\partial\alpha^{*m}\partial\beta^{N-m}\partial\beta^{*N-m}} \delta^{(2)}\left(\alpha\right)\delta^{(2)}\left(\beta\right).
}
 These singularities can be regularized using the methods of Refs. \cite{KieselVogel2010,Agudeloetal2013} to create $P$-functions that are finite everywhere, but there will always be regularized functions that are negative somewhere, which cannot accord with a classical probabilistic description. Highly nonclassical states are at the heart of the states given by Eq. (\ref{eq:general no entanglement generation}) that generate no entanglement.

Within the class of $P$-nonclassical states, we can thus rule out a number of conditions as being sufficient for entanglement generation. $P$-nonclassicality is not a sufficient condition for entanglement generation, nor are nonzero discord and non-Gaussianity. If the states given by Eq. (\ref{eq:general no entanglement generation}) were classical in another way, we could posit that violating this type of classicality would be a necessary condition for entanglement generation; however, no such classical feature has been discovered. We are left with $P$-nonclassicality as the sole necessary condition for entanglement generation at beam splitters.
}

It is intriguing to note that there are many instances in which any SU(2) unpolarized state is operationally equivalent to a pure vacuum state in terms of entanglement generation properties. There may be circumstances in which such an unpolarized state is easier to prepare than the two mode vacuum, using, e.g., a \textblue{polarization-randomized laser}, and will yield the same experimental results as the latter state. Further studies should be done to investigate other states that can mimic the entanglement properties of hard-to-produce states.

\section{Extension to $n$ modes}
In this section we investigate the possibility of finding $n$-mode mixed states that remain separable under the action of linear optics. These operations are described by SU($n$); for pure states, Ref. \cite{Zhangetal2013} showed that the only $n$-mode states that remain separable under connected SU($n$) operators $\mathcal{U}$ are special types of squeezed coherent states.

Our analysis of the two mode case is easily extended to the $n$-mode case, by again noting that the pure states that remain separable form a convex set. Thus, convex combinations of the special pure squeezed states found by Ref. \cite{Zhangetal2013} will also yield zero entanglement at the output of an array of passive linear optical devices. 

In $n$ dimensions, we also have that $\mathcal{U}\sum_N\lambda_N\hat{\mathds{I}}_N\mathcal{U}^\dagger=\sum_N\lambda_N\hat{\mathds{I}}_N$, implying that we may again generalize this convex set along the lines of condition (iii) above. This is a valid procedure because the state 
\begin{equation}
\sum_N \lambda_N \hat{\mathds{I}}_N = \sum_N \sum_{k_1+\cdots+k_n=N} \lambda_N \left| k_1\right\rangle \left\langle k_1\right|\otimes \cdots \otimes \left| k_n\right\rangle \left\langle k_n\right|
\end{equation} is fully separable across all $n$ modes. The generalization by way of SU($n$) unpolarized states does indeed hold for $n>2$, and so we say the $n$-mode states that do not generate entanglement under SU($n$) operations are SU($n$) unpolarized states squeezed by the same amounts along the correct axes dictated by the SU($n$) operation, displacements thereof, and convex combinations of the results (including the maximally mixed state).

\section{Conclusions}
In this paper we have investigated the effects of SU(2) operations on separable states, corresponding to the entangling properties of beam splitters. We quantified the amount of entanglement generated by small rotations, finding that perfectly polarized states generate more entanglement as the number of particles is more evenly distributed between the particle number subspaces. For general two-mode pure states, we found in Eq. (\ref{eq:order theta2 entanglement}) that a similar condition holds for both modes, but that this increase in entanglement can be mitigated by appropriately adjusting the phases of the input states. This corresponded with the known set of nonclassical {pure} states that remain separable after interacting at a beam splitter \cite{Zhangetal2013}. In all cases, entanglement generated grows with the magnitude of the reflectivity of the beam splitter.

We then found {the first} set of {non-Gaussian} mixed states that generate no entanglement at a beam splitter. These are convex combinations of displaced and squeezed unpolarized states, where the squeezing must occur along axes dictated by the beam splitter's axes. {We do not claim to have found a comprehensive list of mixed states that generate no entanglement, due to the lack of a simple general formula to analytically test for mixed state entanglement \cite{Gurvits2004}.} Our result [Eq. (\ref{eq:non classical pure states that do not generate entanglement})] includes the Gaussian states found by Ref. \cite{HuangAgarwal1994}; {moreover, it identifies the first known set of non-Gaussian states with this property}. 

Our set of states can be highly nonclassical in terms of $P$-nonclassicality, in juxtaposition to nonclassicality being a necessary condition for entanglement generation at a beam splitter. Further, our set of states is nonclassical in terms of being one-way quantum-classical correlated, meaning that the information contained within them cannot be fully recovered using local operations and classical communication \cite{Horodeckietal2005}. The earlier known class of pure states only included classical-classical correlated states, which are of little use to quantum information processing (see, however, \textblue{Refs.} \cite{Shahandehetal2017,Shahandehetal2017arxivnewcoherencemeasure}) 

{We further discovered that SU(2) unpolarized states can be operationally equivalent to pure vacuum state in terms of entanglement generation. This leads us to suspect that there are other important entanglement properties of SU(2) unpolarized states waiting to be revealed.}

Finally, we commented on the ability to form sets of states that generate no entanglement under SU($n$) operations. This can be done by taking convex combinations of the states found in Ref. \cite{Zhangetal2013}. Moreover, this set can be extended by using the properties of unpolarized states, as was done in the two mode case.

Our results are important for understanding the entangling properties of beam splitters. We have quantified the entanglement generated by weakly-reflecting beam splitters, and found intriguing sets of states that remain unentangled by all beam splitters. Protocols that generate entanglement via beam splitters, such as boson sampling, must take these effects into account. Our result will further be useful for entanglement characterization and for metrology.

\begin{acknowledgements}
	A.G. would like to thank Nicol\'as Quesada for pointing out important references. This work was supported by the NSERC Discovery Award Fund \#480483 and by the Alexander Graham Bell Scholarship \#504825.
\end{acknowledgements}

\begin{appendix}
\onecolumngrid
\section{Effect of rotation on general two mode state}
\label{app:rotating two mode state}
We here investigate the effect of rotation on an arbitrary, separable two mode state $\ket{\Psi}=\ket{\psi}\otimes\ket{\phi}$. 
We proceed as in Section \ref{sec:perfectly polarized input}:
\eq{
	\hat{\rho}&=\hat{R}\left(\theta,0\right)\ket{\Psi}\bra{\Psi}\hat{R}^\dagger\left(\theta,0\right)\\
	&\approx \ket{\Psi}\bra{\Psi}+\frac{\theta}{2}\left[\left(\ha\hbd-\had\hb\right)\ket{\Psi}\bra{\Psi}+\ket{\Psi}\bra{\Psi}\left(\had\hb-\ha\hbd\right)\right] +\frac{\theta^2}{4}\left(\ha\hbd-\had\hb\right)\ket{\Psi}\bra{\Psi}\left(\had\hb-\ha\hbd\right)\\
	&\quad +\frac{\theta^2}{8}\left[\left(\ha^2\hbd\vphantom{a}^2+\had\vphantom{a}^2\hb^2-2\had\ha\hbd\hb-\had\ha-\hbd\hb\right)\ket{\Psi}\bra{\Psi}+\ket{\Psi}\bra{\Psi}\left(\ha^2\hbd\vphantom{a}^2+\had\vphantom{a}^2\hb^2-2\had\ha\hbd\hb-\had\ha-\hbd\hb\right)\right].
}
Tracing out mode $\hb$ yields
\eq{
	\hat{\rho}_a&\approx \ket{\psi}\bra{\psi}+\frac{\theta}{2}\left(\expct{\hbd}\ha\ket{\psi}\bra{\psi}-\expct{\hb}\had\ket{\psi}\bra{\psi}+\expct{\hb}\ket{\psi}\bra{\psi}\had-\expct{\hbd}\ket{\psi}\bra{\psi}\ha\right)\\
	&\quad+ \frac{\theta^2}{4}\left(\expct{\hb\hbd}\ha\ket{\psi}\bra{\psi}\had+\expct{\hbd\hb}\had\ket{\psi}\bra{\psi}\ha-\expct{\hbd\vphantom{a}^2}\ha\ket{\psi}\bra{\psi}\ha-\expct{\hb^2}\had\ket{\psi}\bra{\psi}\had\right)\\
	&\quad+\frac{\theta^2}{8}\left(\expct{\hbd\vphantom{a}^2}\ha^2\ket{\psi}\bra{\psi} +\expct{\hb^2}\had\vphantom{a}^2\ket{\psi}\bra{\psi}-2\expct{\hbd\hb}\had\ha\ket{\psi}\bra{\psi}-\had\ha\ket{\psi}\bra{\psi}-2\expct{\hbd\hb}\ket{\psi}\bra{\psi} \right.\\
	&\quad\quad\quad\quad\left.+\expct{\hbd\vphantom{a}^2}\ket{\psi}\bra{\psi}\ha^2+\expct{\hb^2}\ket{\psi}\bra{\psi}\had\vphantom{a}^2-2\expct{\hbd\hb}\ket{\psi}\bra{\psi}\had\ha-\ket{\psi}\bra{\psi}\had\ha\right).
}
Then it follows that
\eq{
	\text{Tr}\left[\hat{\rho}_a^2\right]&\approx 1+\left(\frac{\theta}{2}\right)^2\left[2\expct{\hbd}^2\left(\expct{\ha}^2-\expct{\ha^2}\right)+2\expct{\hb}^2\left(\expct{\had}^2
	-\expct{\had\vphantom{a}^2}
	\right)
	+2\expct{\hbd}\expct{\hb}\left(2\expct{\had\ha}+1-2\expct{\ha}\expct{\had}\right)\right]\\
	&\quad+\frac{\theta^2}{4}\left[2\left(2\expct{\hbd\hb}\expct{\ha}\expct{\had}+\expct{\ha}\expct{\had}-\expct{\hbd\vphantom{a}^2}\expct{\ha}^2-\expct{\hb^2}\expct{\had}^2\right)\right]\\
	&\quad+\frac{\theta^2}{8}\left[4\left(\expct{\hbd\vphantom{a}^2}\expct{\ha^2}+\expct{\hb^2}\expct{\had\vphantom{a}^2}-2\expct{\hbd\hb}\expct{\had\ha}-\expct{\had\ha}-\expct{\hbd\hb}
	\right)\right]\\
	&=1-\theta^2\left(AB+\frac{A+B}{2}-\Re\left[\Delta^2\hbd\Delta^2\ha\right]\right)
}
for
\eq{
	A\equiv &\expct{\had\ha}-\expct{\had}\expct{\ha}\\
	B\equiv &\expct{\hbd\hb}-\expct{\hbd}\expct{\hb}
}
and $\Delta^2\hat{O}=\expct{\hat{O}^2}-\expct{\hat{O}}^2$. As before, $A,B\geq 0$, with the equalities holding only for modes $\ha$ and $\hb$ being coherent states. We generalize to the $\varphi\neq0$ case by taking $\ha\to\ha\eu^{\iu\varphi}$ to yield
\eq{
	\text{Tr}\left[\hat{\rho}_a^2\right]=\text{Tr}\left[\hat{\rho}_b^2\right]\approx 1-\theta^2\left(AB+\frac{A+B}{2}-\Re\left[\eu^{2\iu\varphi}\Delta^2\hbd\Delta^2\ha\right]\right).
}

\section{States unaffected by SU(2) operations are separable}
\label{app:unpolarized states separable}
We here show that the states of Eq. (\ref{eq:unpolarized state}) are separable. Expanding Eq. (\ref{eq:unpolarized state}) yields
\eq{
	\hat{\rho}_\text{un}&=\sum_{N=0}^\infty\lambda_N\sum_{m=0}^{N}\ket{m}\bra{m}\otimes\ket{N-m}\bra{N-m}\\
	&=\sum_{m=0}^{\infty}\ket{m}\bra{m}\otimes\sum_{N\geq m}^{}\lambda_N\ket{N-m}\bra{N-m}.
}
We verify that this state is of the form of Eq. (\ref{eq:werner decomposition separable mixed states}) by defining
\eq{
	p_m=1-\sum_{N<m}^{}\lambda_N\left(N+1\right)\geq 0,\quad 0\leq p_m\leq 1
}
and the density operators
\eq{
	\hat{\rho}_m^{(A)}=\ket{m}\bra{m},\quad \hat{\rho}_m^{(B)}=\frac{1}{p_m}\sum_{N\geq m}^{}\lambda_N\ket{N-m}\bra{N-m}.
}
Then $\hat{\rho}_\text{un}=\sum_m p_m\hat{\rho}_m^{(A)}\otimes\hat{\rho}_m^{(B)}$ satisfies all of the properties of a separable state. We note that this type of separable state falls into the category of one-way quantum-classical correlated states, because the states $\left\{\hat{\rho}_m^{(A)}\right\}$ are orthogonal \cite{Horodeckietal2005}.

This class of states includes two-mode thermal states with equal temperatures. Writing $\hat{\rho}^{}_\text{th}\left(T\right)\propto\sum_m\eu^{-m/k_BT} \ket{m}\bra{m}$, we have
\eq{
	\hat{\rho}^{(A)}_\text{th}\left(T\right)\otimes\hat{\rho}^{(B)}_\text{th}\left(T\right)&\propto\sum_{m,n}\eu^{-\frac{m+n}{k_BT}}\ket{m}\bra{m}\otimes\ket{n}\bra{n}\\
	&=\sum_N\sum_{m+n=N}\eu^{-\frac{m+n}{k_BT}}\ket{m}\bra{m}\otimes\ket{n}\bra{n}\\
	&=\sum_N\eu^{-\frac{N}{k_BT}}\sum_{m=0}^N\ket{m}\bra{m}\otimes\ket{N-m}\bra{N-m}.
}
This is clearly of the form of Eq. (\ref{eq:unpolarized state}).

\end{appendix}
\twocolumngrid


%

\end{document}